\documentclass[iop,numberedappendix,appendixfloats]{emulateapj}

\usepackage[breaklinks,colorlinks,citecolor=black,linkcolor=black,urlcolor=black]{hyperref}

\newcommand*{\mysub}[2]{\ensuremath{#1_{\mathrm{#2}}}}
\newcommand*{\unit}[1]{\ensuremath{\mathrm{\, #1}}}

\newcommand*{\Omegam}{\mysub{\Omega}{m}}

\newcommand*{\Omegal}{\mysub{\Omega}{\Lambda}}

\newcommand*{\Lx}{\mysub{L}{X}}
\newcommand*{\Yx}{\mysub{Y}{X}}
\newcommand*{\LCDM}{\ensuremath{\Lambda}CDM}
\newcommand*{\rhocr}{\mysub{\rho}{cr}}
\newcommand*{\sigmaT}{\mysub{\sigma}{T}}
\newcommand*{\me}{\mysub{m}{e}}
\newcommand*{\dA}{\mysub{D}{A}}
\newcommand*{\rs}{\mysub{r}{s}}
\newcommand*{\thetas}{\mysub{\theta}{s}}

\newcommand*{\second}{\unit{s}}

\newcommand*{\erg}{\unit{erg}}
\newcommand*{\cm}{\unit{cm}}
\newcommand*{\km}{\unit{km}}

\newcommand*{\Mpc}{\unit{Mpc}}
\newcommand*{\keV}{\unit{keV}}
\newcommand*{\Msun}{\ensuremath{\, M_{\odot}}}

\newcommand*{\E}[1]{\ensuremath{\times 10^{#1}}}
\newcommand*{\ltsim}{\ {\raise-.75ex\hbox{$\buildrel<\over\sim$}}\ }
\newcommand*{\gtsim}{\ {\raise-.75ex\hbox{$\buildrel>\over\sim$}}\ }
\newcommand*{\proptosim}{\ {\raise-.75ex\hbox{$\buildrel\propto\over\sim$}}\ }

\newcommand*{\secref}{Section}
\newcommand*{\appref}{Appendix}
\newcommand*{\eqnref}{Equation}
\newcommand*{\figref}{Figure}
\newcommand*{\tabref}{Table}

\defcitealias{Sayers1211.1632}{S13}
\newcommand*{\sayers}{\citetalias{Sayers1211.1632}}
\defcitealias{Andersson1006.3068}{A11}
\newcommand*{\Aeleven}{\citetalias{Andersson1006.3068}}
\defcitealias{Willis1212.4185}{W13}
\newcommand*{\willis}{\citetalias{Willis1212.4185}}

\newcommand*{\cl}{XLSSU\,J0217$-$0345}
\newcommand*{\clfull}{XLSSU\,J021744.1$-$034536}

\begin{document}

\submitted{Submitted January 9, 2014; accepted August 25, 2014}

\title{The XXL Survey: V. Detection of the Sunyaev-Zel'dovich Effect of the Redshift 1.9 Galaxy Cluster \clfull{} with CARMA}

\author{A.~B.~Mantz,\altaffilmark{1,2$\star$}
  Z.~Abdulla,\altaffilmark{1,2}
  J.~E.~Carlstrom,\altaffilmark{1,2,3}
  C.~H.~Greer,\altaffilmark{4}
  E.~M.~Leitch,\altaffilmark{1,2}\\
  D.~P.~Marrone,\altaffilmark{4}
  S.~Muchovej,\altaffilmark{5}
  C.~Adami,\altaffilmark{6}
  M.~Birkinshaw,\altaffilmark{7}
  M.~Bremer,\altaffilmark{7}
  N.~Clerc,\altaffilmark{8}\\
  P.~Giles,\altaffilmark{7}
  C.~Horellou,\altaffilmark{9}
  B.~Maughan,\altaffilmark{7}
  F.~Pacaud,\altaffilmark{10}
  M.~Pierre,\altaffilmark{11}
  J.~Willis\altaffilmark{12}
}

\altaffiltext{1}{Department of Astronomy and Astrophysics, University of Chicago, 5640 South Ellis Avenue, Chicago, IL 60637, USA}
\altaffiltext{2}{Kavli Institute for Cosmological Physics, University of Chicago, 5640 South Ellis Avenue, Chicago, IL 60637, USA}
\altaffiltext{3}{Department of Physics/Enrico Fermi Institute, University of Chicago, 5640 South Ellis Avenue, Chicago, IL 60637, USA}
\altaffiltext{4}{Steward Observatory, University of Arizona, 933 North Cherry Avenue, Tucson, AZ 85721, USA}
\altaffiltext{5}{Owens Valley Radio Observatory, California Institute of Technology, Big Pine, CA 93513, USA}
\altaffiltext{6}{LAM, OAMP, Universit\'e Aix-Marseille \& CNRS, P\^ole de l'\'Etoile, Site de Ch\^ateau Gombert, 38 rue Fr\'ed\'eric Joliot-Curie, F-13388, Marseille 13 Cedex, France}
\altaffiltext{7}{H. H. Wills Physics Laboratory, University of Bristol, Tyndall Avenue, Bristol BS8 1TL, UK}
\altaffiltext{8}{Max-Planck Institut f\"ur Extraterrestrische Physik, Giessenbachstrasse 1, D-85748 Garching, Germany}
\altaffiltext{9}{Department of Earth and Space Sciences, Chalmers University of Technology, Onsala Space Observatory, SE-439 92 Onsala, Sweden}
\altaffiltext{10}{Argelander-Institute for Astronomy, Auf dem H\"ugel 71, D-53121 Bonn, Germany}
\altaffiltext{11}{Service d'Astrophysique, Bt. 709, CEA Saclay, F-91191 Gif sur Yvette Cedex, France}
\altaffiltext{12}{Department of Physics and Astronomy, University of Victoria, 3800 Finnerty Road, Victoria, BC, Canada}
\altaffiltext{$\star$}{E-mail: \href{mailto:amantz@kicp.uchicago.edu}{\tt amantz@kicp.uchicago.edu}}

\shorttitle{SZ Effect of a $z=1.9$ Galaxy Cluster}
\shortauthors{A. B. Mantz et al.}

\begin{abstract}
  We report the detection of the Sunyaev-Zel'dovich (SZ) effect of galaxy cluster \clfull{}, using 30\,GHz CARMA data. This cluster was discovered via its extended X-ray emission in the XMM-{\it Newton} Large Scale Structure survey, the precursor to the XXL survey. It has a photometrically determined redshift $z=1.91^{+0.19}_{-0.21}$, making it among the most distant clusters known, and nominally the most distant for which the SZ effect has been measured. The spherically integrated Comptonization is $Y_{500}=(3.0\pm0.4)\E{-12}$, a measurement which is relatively insensitive to assumptions regarding the size and redshift of the cluster, as well as the background cosmology. Using a variety of locally calibrated cluster scaling relations extrapolated to $z\sim2$, we estimate a mass $M_{500} \sim (1$--$2)\E{14}\Msun$ from the X-ray flux and SZ signal. The measured properties of this cluster are in good agreement with the extrapolation of an X-ray luminosity--SZ effect scaling relation calibrated from clusters discovered by the South Pole Telescope at higher masses and lower redshifts. The full XXL-CARMA sample will provide a more complete, multi-wavelength census of distant clusters in order to robustly extend the calibration of cluster scaling relations to these high redshifts.
\end{abstract}


\section{Introduction} \label{sec:intro}

\setcounter{footnote}{12} % cf number of affiliations

Building on the success of cosmological tests using the number density and growth of galaxy clusters (e.g., \citealt{Mantz0709.4294, Mantz0909.3098, Vikhlinin0812.2720, Rozo0902.3702, Benson1112.5435, Hasselfield1301.0816}) and cluster gas mass fractions (e.g., \citealt{Allen0405340, Allen0706.0033, LaRoque0604039, Ettori0904.2740, Mantz1402.6212}; see also \citealt*{Allen1103.4829}), a number of observational programs seek to extend the census of the cluster population to redshifts $z\ge1$. Efforts have included searches for serendipitous detections \citep{Fassbender1111.0009, Mehrtens1106.3056}, and controlled surveys \citep{Eisenhardt0804.4798, Muzzin0807.0227, Hasselfield1301.0816, Planck1303.5089, Reichardt1203.5775}. The number of confirmed, high-redshift clusters has recently expanded rapidly, including $z\gtsim1.5$ discoveries at X-ray \citep{Fassbender1101.3313, Santos1105.5877}, millimeter \citep{Bayliss1307.2903} and IR \citep{Papovich1002.3158, Gobat1011.1837, Brodwin1205.3787, Stanford1205.3786, Zeimann1207.4793} wavelengths.

This paper concerns a cluster discovered via its extended X-ray emission in the XMM-{\it Newton} Large Scale Structure  survey  (XMM-LSS; \citealt{Pierre0305191}). XMM-LSS reaches a flux detection limit for clusters of $\sim5\times10^{-15}\erg\cm^{-2}\second^{-1}$ in the 0.5--2.0\,keV band over an 11 sq.\ deg.\ footprint that has extensive, complementary optical and IR photometry. Clusters discovered by the survey reach redshifts as high as $z\sim2$ \citep[][hereafter \willis{}; see also \citealt{Valtchanov0305192, Willis2005MNRAS.363..675, Bremer0607425, Pierre0607170, Pacaud0709.1950, Maughan0709.2300}]{Willis1212.4185}. An expanded survey footprint covering 50 sq.\ deg.\ (XXL) has since been completed to similar depth, and a number of cosmological and astrophysical investigations based on these data are ongoing \citep{Pierre1009.3182}.

Among these investigations is an observing campaign with the Combined Array for Research in Millimeter-wave Astronomy\footnote{\url{http://www.mmarray.org}} (CARMA), targeting cluster detections in the northern 25 sq.\ deg.\ field of XXL (which includes the XMM-LSS footprint) at 30\,GHz, with the aim of measuring the Sunyaev-Zel'dovich (SZ) effect of the hot intracluster medium (ICM). Here we present the first result from that program, the detection of the SZ signal of cluster \clfull{} (hereafter \cl{}), which was obtained using the CARMA sub-array of eight 3.5\,m telescopes (formerly the SZA). This cluster is a ``class 1'' detection, meaning  that it meets conservative criteria designed to produce a pure sample of extended sources (see \citealt{Pacaud0607177}). The survey data imply an unabsorbed 0.5--2.0\,keV flux of $1.08\E{-14}\erg\cm^{-2}\second^{-1}$ (\willis{}). With a photometric redshift $\sim1.9$ (see below) and only $\sim100$ net source photons contributing to the detection, this cluster is representative of a population of faint, distant X-ray sources detectable by XXL (and by upcoming missions like eROSITA; \citealt{Predehl1001.2502}) for which survey data cannot directly provide estimates of the temperature or mass of the ICM. Observations of the SZ effect provide a complementary probe, allowing independent confirmation of the presence of hot gas and estimates of the cluster mass. The redshift independence of the SZ brightness makes it particularly powerful for following up distant clusters.

\willis{} present 10-band ($ugrizYJK,3.6\micron,4.5\micron$) photometric data covering \cl{}, and we summarize the key results of that analysis here. The distribution of photometric redshifts for galaxies near the X-ray detection is bimodal, with a peak at $z\sim1.0$ and another (larger) peak at $z\sim1.9$. Galaxies associated with the $z\sim1$ peak display neither spatial clustering associated with the X-ray source nor an identifiable red sequence. In contrast, galaxies with photometric redshifts $1.7<z<2.1$ are clustered around the location of the extended X-ray emission (\figref~\ref{fig:map}) and display a poorly populated yet significant red sequence, whose color is consistent with that anticipated from a passively evolving, solar metallicity stellar population formed at $z=10$ and observed at $z=1.9$. On this basis, \willis{} conclude that the $z\sim1$ peak arises from an unassociated foreground structure, and assign a cluster redshift $z=1.91^{+0.19}_{-0.21}$. Note that the quoted uncertainty represents the full width of the peak in the galaxy redshift histogram, not the error on the mean of this peak. Spectroscopic confirmation of this redshift is not yet available, though we note that comparison of the cluster's X-ray flux with our CARMA data favors a redshift $\sim1.9$ compared with $1.0$ (see \secref~\ref{sec:LYscaling}).

This paper is organized as follows. \secref{}s~\ref{sec:data} and \ref{sec:analysis} describe the CARMA data and the determination of the cluster SZ signal, the spherically integrated Compton $Y$ parameter. In \secref~\ref{sec:discussion}, we discuss estimates of the cluster mass based on the X-ray and SZ data, and compare the measured X-ray and SZ signals for \cl{} to a scaling relation calibrated from higher-mass and lower-redshift South Pole Telescope (SPT) clusters. We summarize in \secref~\ref{sec:conclusion}.

Throughout this work, we assume a concordance \LCDM{} cosmological model, with dark energy in the form of a cosmological constant, described by Hubble parameter $h_{70}=H_0/70\km\second^{-1}\Mpc^{-1}=1$, matter density $\Omegam=0.3$ and dark energy density $\Omegal=0.7$. Quoted uncertainties refer to 68.3\% confidence intervals (with the exception of the photometric redshift estimate, noted above). We report dimensionless, spherically integrated Comptonization ($Y$) in units of steradians.

\newpage
\section{CARMA Data} \label{sec:data}

CARMA is a heterogeneous interferometric array comprised of six 10.4\,m, nine 6.1\,m, and eight 3.5\,m telescopes; our data were obtained using the eight-element array of 3.5\,m antennas operating at a central frequency of 31\,GHz. The data were taken over two periods spanning 2012 March--July and 2013 September--November. The 3.5\,m antennas were configured with six elements in a compact array, providing 15 baselines with sensitivity at arcminute scales, and two outlying elements providing 13 baselines with sensitivity at higher resolution. The compact and extended baselines respectively sample {\it uv} ranges of 0.35--2\,k$\lambda$ and 2--9.5\,k$\lambda$ with comparable flux sensitivity, allowing emission from compact radio sources to be distinguished from the extended SZ effect. The signals were processed by the CARMA 8\,GHz bandwidth digital correlator in sixteen 500\,MHz sub-bands, each consisting of 16 channels. More details of the specific observations can be found in \tabref~\ref{tab:obs}.

\begin{table}
  \begin{center}
    \caption{CARMA Data}
    \label{tab:obs}
    \vspace{1ex}
    \begin{tabular}{lccc}
      \hline
      \multicolumn{1}{c}{ID} & UT date & config\tablenotemark{a} & $t_\mathrm{int}$\tablenotemark{b} \\
      \hline
      c0927.3SL\_31J02170.1  &  2012-03-06  &  SL  & 2.5 \\
      c0927.3SL\_31J02170.2  &  2012-03-09  &  SL  & 3.5 \\
      c0927.3SL\_31J02170.3  &  2012-03-10  &  SL  & 3.5 \\
      c0927.3SL\_31J02170.4  &  2012-03-12  &  SL  & 2.2 \\
      c0927.3SL\_31J02170.5  &  2012-03-13  &  SL  & 2.6 \\
      c0927.3SL\_31J02170.6  &  2012-03-14  &  SL  & 2.3 \\
      c0927.3SL\_31J02170.7  &  2012-03-15  &  SL  & 1.6 \\
      c0927.3SH\_31J02170.1  &  2012-05-31  &  SH  & 3.6 \\
      c0927.3SH\_31J02170.3  &  2012-06-04  &  SH  & 2.5 \\
      c0927.3SH\_31J02170.4  &  2012-07-24  &  SH  & 1.9 \\
      c0927.3SH\_31J02170.5  &  2012-09-01  &  SH  & 2.2 \\
      c1171.33SH\_30s4.3     &  2013-11-01  &  SH  & 0.4 \\
      c1171.33SH\_30s4.4     &  2013-11-02  &  SH  & 3.5 \\
      c1171.33SH\_30s4.5     &  2013-11-03  &  SH  & 3.5 \\
      c1171.33SH\_30s4.7     &  2013-11-04  &  SH  & 3.1 \\
      c1171.33SH\_30s4.8     &  2013-11-05  &  SH  & 3.2 \\
      c1171.33SH\_30s4.9     &  2013-11-06  &  SH  & 2.8 \\
      c1171.33SH\_30s4.10    &  2013-11-08  &  SH  & 0.1 \\
      c1171V.35SH\_30s6.7    &  2013-11-10  &  SH  & 1.7 \\
      c1171.33SH\_30s4.11    &  2013-11-11  &  SH  & 3.1 \\
      c1171V.36SH\_30s7.2    &  2013-11-12  &  SH  & 3.0 \\
     \hline
    \end{tabular}
  \end{center}
  \tablenotetext{1}{Configuration of the 3.5\,m CARMA dishes. The SH and SL configurations provide similar baseline coverage, and are approximately equivalent for the targets at the declination of \cl{}.}
  \tablenotetext{2}{Effective on-source integration time after flagging (hours).}
\end{table}

\begin{figure*}
  \centering
  \includegraphics{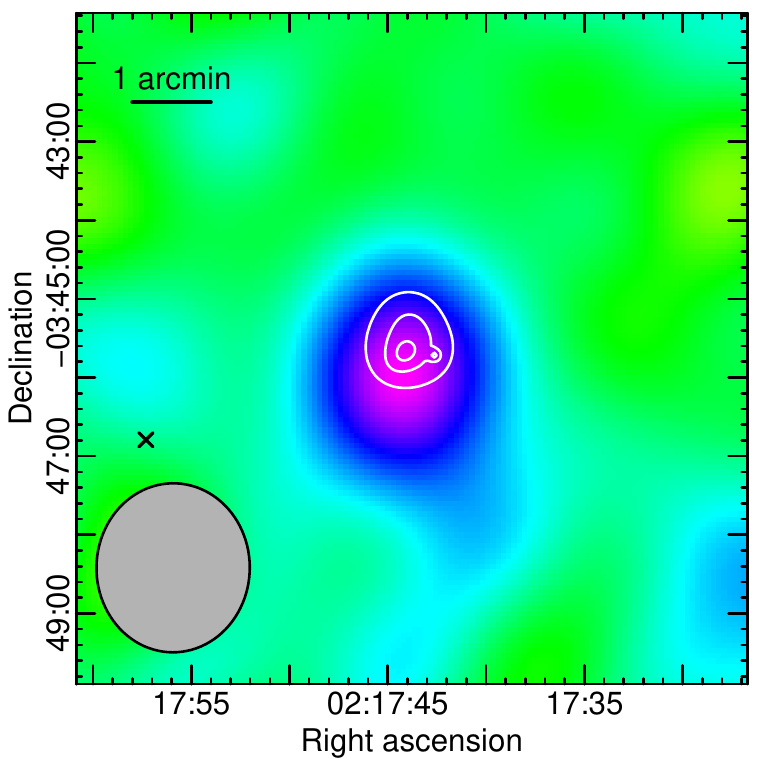}
  \hspace{-3mm}
  \includegraphics{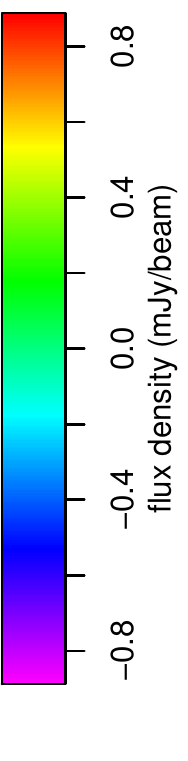}
  \hspace{5mm}
  \includegraphics[scale=0.23]{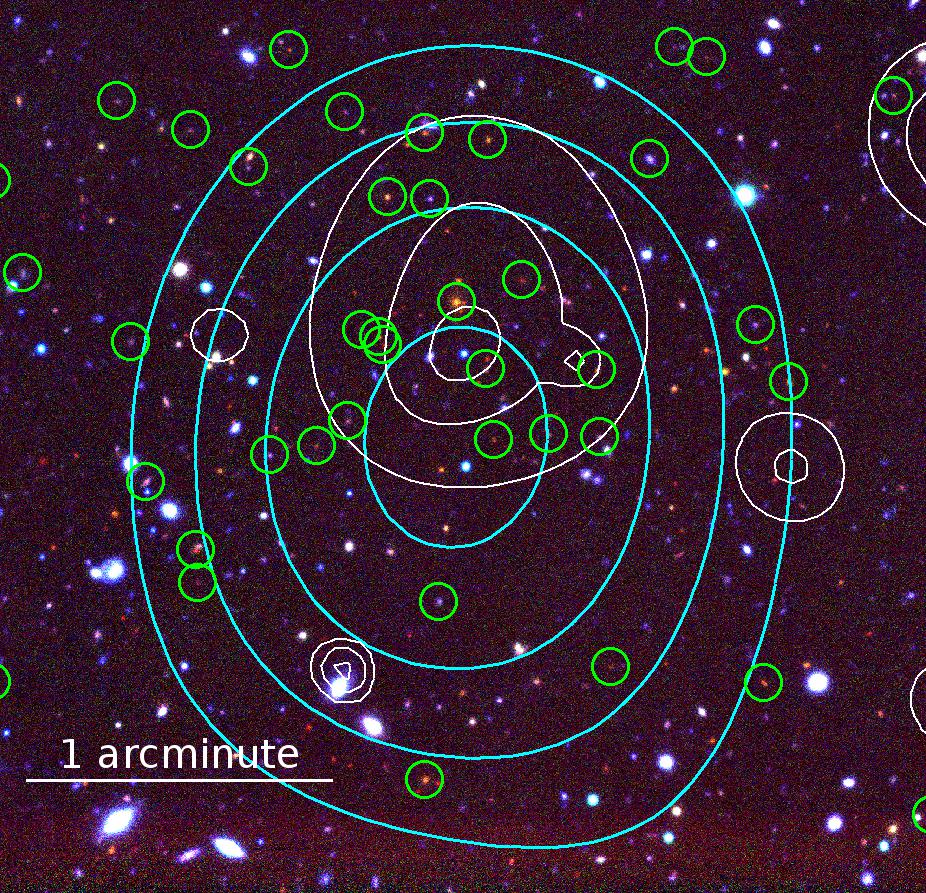}
  \caption{
    Left: Short-baseline ($uv$ radii $<2$k$\lambda$) 30\,GHz map of \cl{} after modeling and subtracting point sources and applying the CLEAN algorithm. The position of the brighter point source in \tabref~\ref{tab:ptsrc} is indicated by the ``$\times$'' (the other lies outside the image). White contours show the extended X-ray emission associated with the cluster detection. The gray ellipse in the lower-left corner shows the FWHM synthesized beam. Right: $iJK$ image, with X-ray (white) and SZ (blue) contours overlaid. The SZ contours correspond to $-2.5$, $-3.5$, $-4.5$, and $-5.5$ times the rms noise level of the short-baseline map. Galaxies with photometric redshifts in the range $1.7<z<2.1$ are circled in green.
  }
  \label{fig:map}
\end{figure*}
 
Our data reduction procedure is described in \citet{Muchovej0610115}. Briefly, the data are filtered for bad weather, shadowing, and high system temperatures or other technical issues, and bandpass and gain calibrations are applied based on periodic observations of planets and radio-bright quasars. The absolute flux calibration compares observations of Mars to the model of \citet{Rudy1987PhDT.........4R}, which is accurate to better than 5\%. The reduced, calibrated data are equivalent to a total of 52.8 hours of on-source integration time.  At short baselines ($uv$ radii $<2$\,k$\lambda$) the rms noise level is 0.15\,mJy$/$beam, with a $117\times129$ arcsec full-width-at-half-maximum (FWHM) synthesized beam. At long baselines ($>2$\,k$\lambda$) , the noise is 0.13\,mJy$/$beam, and the synthesized beam FWHM is $17.9\times24.7$ arcsec.

The short-baseline map, after modeling and removing point sources (\secref~\ref{sec:analysis}) and applying the CLEAN image deconvolution algorithm of \citet{Hoegbom1974A26AS...15..417}, is shown in the left panel of \figref~\ref{fig:map}. (Note that in our analysis of the cluster, detailed in the next section, we do not use this map, but instead fit models directly to the measured visibility data in the $uv$ plane.) The right panel of the figure shows an $iJK$ color image of the cluster from \willis{}, with X-ray brightness and SZ significance contours overlaid, and galaxies with photometric redshifts between 1.7 and 2.1 circled.

\section{Measurements of Comptonization and Mass} \label{sec:analysis}

We use a Markov Chain Monte Carlo exploration of the parameter space to fit models of the cluster SZ effect and emission from unresolved radio sources directly to the $uv$ data at all baselines. To parametrize the SZ signal, we adopt the \citet{Nagai0703661} form for the shape of the three-dimensional, spherically symmetric electron pressure profile of clusters,
\begin{equation} \label{eq:gnfw}
P(x) = \frac{P_0}{x^\gamma \left[1+x^\alpha\right]^{(\beta-\gamma)/\alpha}},
\end{equation}
and fix the parameters $\alpha=0.86$, $\beta=3.67$ and $\gamma=0.67$ as prescribed by \citet[][hereafter \sayers{}]{Sayers1211.1632}. We discuss the implications of using this particular measurement of the parameters in \eqnref~\ref{eq:gnfw} in Appendix~\ref{sec:templates}. The free parameters of this model are the position of the cluster center; the scale radius, which normalizes the radial coordinate ($x=r/\rs$, or $\theta/\thetas$ in units of angle); and an overall normalization, which we take to be the line-of-sight Comptonization through the cluster center,
\begin{equation} \label{eq:y0}
  y_0 = \frac{2\sigmaT\,\dA(z)}{\me c^2} \int_{0}^{\infty} \, d\theta \, P(\theta/\mysub{\theta}{s}),
\end{equation}
where \sigmaT{} is the Thompson cross section, \me{} is the electron rest mass, $c$ is the speed of light in vacuum, and $\dA(z)$ is the angular diameter distance to the cluster.\footnote{Note that measuring $y_0$ or $Y$ does not require explicit assumptions about the cluster redshift or the cosmic expansion history. The factors of $\dA(z)$ appearing in \eqnref{}s~\ref{eq:y0}--\ref{eq:Y} are only necessary to express these quantities in terms of the electron pressure profile.}
The Comptonization integrated in a sphere corresponding to angular radius $\Theta$ is
\begin{equation} \label{eq:Y}
  Y(\Theta) = \frac{\sigmaT\,\dA(z)}{\me c^2} \int_0^{\Theta} 4\pi \, \theta^2 \, d\theta \, P(\theta/\mysub{\theta}{s}).
\end{equation}
Given a prescription for determining $r_{500}$,\footnote{Defined as the cluster radius enclosing a mean density 500 times the critical density of the Universe at the cluster's redshift: $M_{500}=(2000/3)\pi\rhocr(z)r_{500}^3$.} we thus can straightforwardly calculate $Y_{500} = Y[r_{500}/\dA(z)]$. The closely related quantity $\dA(z)^2 \,Y_{500}$ is proportional to the thermal energy of the ICM, and should therefore scale with the total cluster mass.

The measured properties of unresolved radio sources in the field are essentially independent of the cluster gas model because they are  primarily constrained by the long baseline data, where signal from the cluster is negligible ($uv$ radii $\gtsim2$\,k$\lambda$). There are two compact, emissive sources detected in the data, both several arcmin from the cluster position, and we fit for their positions and flux densities simultaneously with the cluster model. The results appear in \tabref~\ref{tab:ptsrc}; we find flux densities at 30\,GHz of $2.21\pm0.09$ and $0.70\pm0.14$\,mJy. The brighter of the two has a 1.4\,GHz counterpart in the NVSS and FIRST surveys and is also the closest 1.4\,GHz source to the cluster position (i.e., there are no radio sources in projection with the cluster SZ decrement in \figref~\ref{fig:map}). There is negligible covariance between the brightness of these sources and cluster parameters in our fits. In particular, a factor of $\sim2$ change to the flux density of the brighter and closer point source would be required to force a $1\sigma$ shift in the normalization of the cluster model.

\begin{table}
  \begin{center}
    \caption{Properties of Unresolved Radio Sources}
    \label{tab:ptsrc}
    \vspace{1ex}
    \begin{tabular}{l@{ $\pm$ }ll@{ $\pm$ }lcc}
      \hline
      \multicolumn{2}{c}{RA} & \multicolumn{2}{c}{Dec} & Offset & flux (mJy) \\
      \hline
      02:17:57.30 & $0.6''$ & $-$03:46:47.5 & $0.6''$ & 3.55$'$ & $2.21\pm0.09$ \\
      02:17:51.84 & $2.5''$ & $-$03:40:29.4 & $2.5''$ & 5.48$'$ & $0.70\pm0.14$ \\
      \hline
    \end{tabular}
  \end{center}
  \tablecomments{J2000 positions, angular distances from the XXL cluster position (02:17:43.9$-$03:45:36), and 30\,GHz flux densities (corrected for the primary beam) of unresolved radio sources detected near \cl{} in our observations.}
\end{table}

The scale radius of the cluster pressure profile is poorly constrained by our data, leading to a strong degeneracy between the scale radius and the normalization of the profile. In the following sections, we consider three priors on the scale radius. The first is based on a scaling relation between mass and pressure, and directly links the scale radius with the measured SZ signal. The second is a Gaussian prior based on the measured X-ray flux from the cluster, and is independent of the SZ data. The third case is simply a wide uniform prior, encompassing values consistent with the other two approaches. In all cases, the addition of the cluster model yields a significantly improved fit compared to a point-source-only model, with $\Delta\chi^2_\mathrm{min} \sim 85$ (for 55\,005 degrees of freedom in the case of \secref~\ref{sec:a10mass}, equivalent to a $7.3\sigma$ significance). Statistically, the cluster models in the following sections provide equally good fits to the data. Their constraints on the integrated Comptonization, $Y_{500}$, agree within measurement uncertainties. These analyses also produce estimates of the cluster mass, which are more sensitive to a priori assumptions than the $Y_{500}$ constraints; we will discuss the mass estimates further in \secref~\ref{sec:mass}. \tabref~\ref{tab:szfit} summarizes the constraints on cluster parameters.

We note an offset of $34\pm9$ arcsec between the nominal X-ray detection position and the best-fitting centers of the SZ models below, although the SZ centers still lie within the extended X-ray emission and the distribution of photometrically selected $z\sim1.9$ galaxies (\figref~\ref{fig:map}). While the X-ray/SZ offset is suggestive of an asymmetry in the distribution of hot gas in the system, we have no reliable information about the morphological state of this cluster and firm conclusions cannot be drawn from the data in hand. Such offsets are not unprecedented \citep{Andersson1006.3068}, nor necessarily unusual in the case of a major merger \citep{Zhang1406.4019}.

\subsection{Scale Radius from the Pressure--Mass Relation} \label{sec:a10mass}

We first consider a cluster model where the scale radius of the pressure profile is self-consistently related to its normalization and the cluster mass through a scaling relation linking pressure and mass. In this case, the single free parameter describing the pressure profile of the cluster is $M_{500}$, which determines the scale radius through the fixed parameter $c_{500}=r_{500}/\rs= 1.18$ (\sayers{}). The normalization of the electron pressure profile is given by
\begin{eqnarray}
  P_0(M_{500}) &=& \left(0.0158\frac{\mathrm{keV}}{\mathrm{cm}^3}\right) E(z)^{8/3} \, h_{70}^{1/2} \hspace{2cm}\nonumber\\
  &&\times \left(\frac{M_{500}}{10^{15}h_{70}^{-1}\Msun}\right)^{\frac{2}{3} + \alpha_P},
\end{eqnarray}
with $\alpha_P=0.12$, and where we have neglected a higher-order correction in the exponent of $M_{500}$ (see \citealt{Arnaud0910.1234} and \sayers{}). Here $E(z)=H(z)/H_0$ encodes the evolution of the Hubble parameter.

This analysis yields a mass estimate of $M_{500}=(1.34\pm0.11)\E{14}\Msun$, corresponding to scale radius $\rs=0.329\pm0.009\Mpc$ (statistical uncertainties only). Note that we have simply extrapolated the \sayers{} scaling relation, calibrated at $z<0.9$, to $z=1.9$. This $M_{500}$ constraint, as with any such extrapolated estimate, should therefore be treated with extreme caution. The constraint on the spherically integrated Comptonization is $Y_{500}=(2.8\pm0.4)\E{-12}$, consistent with the results of the less constrained analyses in the following sections.

\begin{table}
  \begin{center}
    \caption{Cluster Properties}
    \label{tab:szfit}
    \vspace{1ex}
    \begin{tabular}{l@{\hspace{0.5ex}}r@{ $\pm$ }lr@{ $\pm$ }lr@{ $\pm$ }l}
      \hline
      & \multicolumn{2}{c}{\secref~\ref{sec:a10mass}} & \multicolumn{2}{c}{\secref~\ref{sec:xprior}} & \multicolumn{2}{c}{\secref~\ref{sec:uprior}} \\
      \hline
      Offset E ($''$)& $3$&$7$ & $2$&$7$ & $3$&$8$\vspace{0.5ex}\\
      Offset N ($''$) & $-35$&$9$ & $-34$&$9$ & $-34$&$10$\vspace{0.5ex}\\
      $10^4\,y_0$ & $2.5$&$0.2$ & $3.5$&$0.6$ & \multicolumn{2}{c}{$2^{+4}_{-1}$}\vspace{0.5ex}\\
      $\thetas$ ($'$) & $0.652$&$0.017$ & $0.58$&$0.07$ & \multicolumn{2}{c}{$0.4^{+0.4}_{-0.2}$}\vspace{2ex}\\
      $10^{12}\,Y_{500}$ & $2.8$&$0.4$ & $3.0$&$0.4$ & \multicolumn{2}{c}{$2.2^{+1.8}_{-0.8}$}\vspace{0.5ex}\\
      $\dA^2 Y_{500}$ ($10^{-6}$\,Mpc$^2$) & $8.3$&$1.2$ & $9.1$&$1.3$ & \multicolumn{2}{c}{$6^{+5}_{-2}$}\vspace{0.5ex}\\
      $r_{500}$ (Mpc) & $0.388$&$0.010$ & $0.35$&$0.04$ & \multicolumn{2}{c}{---}\vspace{0.5ex}\\
      $M_{500}$ ($10^{14}M_\odot$) & $1.34$&$0.11$ & $1.0$&$0.4$ & \multicolumn{2}{c}{---} \\
      \hline
    \end{tabular}
  \end{center}
  \tablecomments{Best fitting values and 68.3\% confidence intervals for the cluster parameters of our SZ model. Offsets describe the position of the cluster model center with respect to 02:17:43.9$-$03:45:36 in J2000 coordinates, and $y_0$ is the dimensionless line-of-sight Comptonization through the cluster center. Note that there are different priors used in each set of results. In particular, the model of \secref{}~\ref{sec:a10mass} deterministically links $y_0$ and $\thetas$, resulting in tight constraints on those parameters, as well as on the mass. In \secref~\ref{sec:xprior} we instead apply a Gaussian prior to $\thetas$, but do not impose a prior on $y_0$. In \secref~\ref{sec:uprior}, we use a uniform prior on \thetas{}. The basis of the mass constraints also differs; see the indicated sections for details. The $\dA^2 Y_{500}$, $r_{500}$ and $M_{500}$ constraints assume a cluster redshift $z=1.91$.}
\end{table}

\subsection{Scale Radius from the X-ray Flux} \label{sec:xprior}

The XMM measurement of the X-ray luminosity of \cl{} can be used to define a loose prior on the pressure scale radius, providing an approach to fitting the SZ data which is less constraining than the one in \secref~\ref{sec:a10mass}. This procedure introduces a covariance between the measured values of \Lx{} and $Y$. In practice, however, this covariance is negligible compared to the measurement uncertainties, as shown below.

\begin{figure*}
  \centering
  \includegraphics{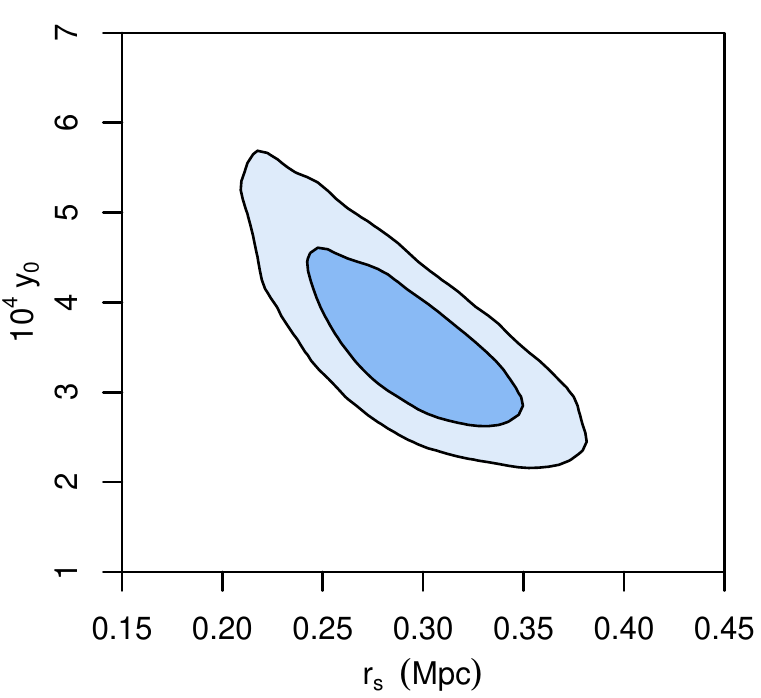}
  \hspace{10mm}
  \includegraphics{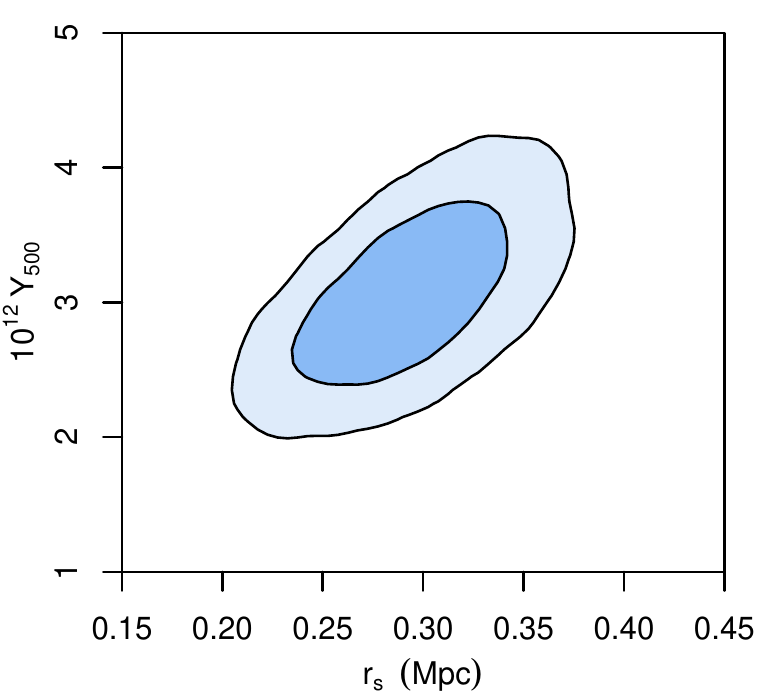}
  \caption{
    Two-dimensional 68.3\% and 95.4\% confidence regions from our analysis in \secref~\ref{sec:xprior}. Left: scale radius and normalization of the pressure profile, parametrized by the line-of-sight Comptonization through the cluster center. Right: scale radius versus the spherically integrated Comptonization within $r_{500}$.
  }
  \label{fig:contours}
\end{figure*}

The \citet{Kaiser1986MNRAS.222..323K} self-similar model relating X-ray luminosity and mass is
\begin{equation} \label{eq:LMscaling}
  \frac{L_\Delta}{E(z)} \propto \left[ E(z) \, M_\Delta \right]^a,
\end{equation}
with $a=4/3$ for the bolometric luminosity, and where both quantities are integrated within a fixed overdensity with respect to critical (e.g., $\Delta=500$). The factors of $E(z)$ arise from the critical density, $\rhocr(z) \propto H(z)^2$. For soft-band emission from clusters (such as the 0.5--2.0\keV{} band used for XXL), measured values of $a$ span the range $\sim1.3$--$1.8$ (e.g., \citealt{Reiprich0111285, Allen0208394, Pratt0809.3784, Vikhlinin0805.2207, Mantz0909.3099}). For convenience, and because the precise value within this range has a negligible effect on our results ($<1$ arcsec in the estimate of $\thetas$), we take $a=4/3$. Using that $M_\Delta \propto E(z)^2\,r_\Delta^3$, the above relation thus yields
\begin{equation} \label{eq:rle}
  r_\Delta \propto L_\Delta^{\frac{1}{3a}} \, E(z)^{-\frac{3a+1}{3a}} \propto L_\Delta^{1/4} \, E(z)^{-5/4}.
\end{equation}

We calibrate the normalization of this relation using other XXL clusters, spanning redshifts $0.4<z<1.1$, for which deep X-ray observations provide estimates of mass and $r_{500}$ (Giles et~al., in preparation).\footnote{The calibration uses 5 clusters in the XMM-XXL sample with comparable X-ray luminosity to \cl{} (also detected by CARMA, to be reported in future work). Luminosities are estimated directly from the survey count rates, and masses from measurements of $\Yx$ (the product of gas mass and temperature) from either the survey data or deeper XMM data, assuming a reference $\Yx$--mass relation. More details, as well as a more complete analysis of cluster scaling relations, will appear in Giles et~al.\ (in prep). Note that the data used here are not powerful enough to directly provide a meaningful constraint on the slope, $a$.} Identifying the 0.5--2.0\,keV \Lx{} values estimated from the survey data with $L_{500}$ for this purpose, we find
\begin{equation} \label{eq:rLscaling}
  \frac{r_{500}}{\mathrm{Mpc}} = (1.035\pm0.021) \,  \left(\frac{\Lx}{10^{44}\erg\second^{-1}}\right)^{1/4} \, E(z)^{-5/4},
\end{equation}
with a residual scatter of $\sim4$\%.

Owing to the use of the luminosity--mass scaling in \eqnref~\ref{eq:LMscaling}, we expect an additional uncertainty of $\sim11$\% to apply to estimates of $r_{500}$, given an intrinsic scatter in $\Lx$ at fixed $M_{500}$ of $40$--$45$\% and an \Lx--$M_{500}$ slope of $\sim4/3$. In practice, we identify $\rs=r_{500}/c_{500}$, with $c_{500}=1.18$ from \sayers{} and $r_{500}$ estimated from \eqnref~\ref{eq:rLscaling}, and marginalize over a Gaussian prior with a standard deviation of 12\%, accounting for the systematic and statistical sources of uncertainty. Assuming the nominal redshift of 1.91, this yields $\rs=0.30\pm0.04\Mpc$, implying a mass $M_{500}=(1.0\pm0.4)\E{14}\Msun$.

Marginalizing over this \Lx{}--motivated prior on the pressure scale radius, we constrain the normalization and center of the cluster model from the SZ data. Our constraints from this analysis are summarized in \tabref~\ref{tab:szfit}; we find $Y_{500}=(3.0\pm0.4)\E{-12}$, or equivalently $\dA(z)^2\,Y_{500}=(9.1\pm1.3)\E{-6}\Mpc^2$.

As described above, there is a strong degeneracy between the scale radius and the pressure normalization, which can be seen in the joint constraints on these parameters shown in the left panel of \figref~\ref{fig:contours}. As the right panel of the figure shows, the scale radius is also degenerate with the spherically integrated Comptonization, $Y_{500}$ (with a somewhat weaker degree of correlation). Despite this degeneracy, however, our results are not particularly sensitive to the cluster redshift given an observationally motivated prior such as the one used in this section. Varying the cluster redshift by $\pm0.2$, and adjusting the prior on \thetas{} accordingly, we find that $Y_{500}$ changes by only $\mp4\%$, well within our statistical uncertainties.

While we performed the test above by repeating the entire analysis assuming a different value for the cluster redshift, an approximate estimate for the size of the effect can be obtained as follows. Taking advantage of the fact that $Y_{500}\proptosim\rs^1\propto r_{500}$ in \figref~\ref{fig:contours}, the fractional shift in $Y_{500}$ with redshift follows from converting \eqnref~\ref{eq:rLscaling} into an expression for \thetas{} and accounting straightforwardly for the luminosity-distance dependence of \Lx{}. From this, we estimate that differences of $>0.5$ in redshift would be necessary to change the measured value of $Y_{500}$ by an amount comparable to its statistical uncertainty ($\sim14\%$). Note that this shift in $Y_{500}$ partially cancels the explicitly redshift-dependent factor $E(z)\dA(z)^2$ appearing in the scaling relation analysis of \secref~\ref{sec:LYscaling}.

\subsection{Uniform Prior on the Scale Radius} \label{sec:uprior}

Without any external constraint, our data are unable to meaningfully break the degeneracy between the scale radius and normalization of the pressure profile, leading to significantly less tight constraints compared to the previous sections (\tabref~\ref{tab:szfit}). However, it is worth noting that the scale radius constraint in this case is consistent with the results of \secref{}s~\ref{sec:a10mass} and \ref{sec:xprior}, rather than preferring extremely low or high values, and that the constraints on the cluster position are comparable.

\section{Discussion} \label{sec:discussion}

\subsection{Mass Estimates} \label{sec:mass}

Our analysis has produced two estimates of the mass of \cl{}: $M_{500}\sim1.3\E{14}\Msun$ from the SZ data in combination with the pressure scaling relation of \sayers{} (\secref~\ref{sec:a10mass}), and $M_{500}\sim1.0\E{14}\Msun$ from the preliminary, empirical \Lx{} scaling for XXL clusters specifically (\secref~\ref{sec:xprior}; Giles et~al. in preparation). We stress that any such estimates based on the extrapolation of scaling relations beyond the mass and redshift regimes where they are calibrated must be regarded skeptically. Nevertheless, for completeness, we collect in \tabref~\ref{tab:masses} a few more mass estimates based on the adopted cluster redshift of $z=1.91$ and either the survey X-ray luminosity or our measurement of $Y_{500}$ from \secref~\ref{sec:xprior}. Given the large and poorly quantified systematic uncertainties associated with the extrapolation in redshift, we do not quote errors for these estimates. Nevertheless, the overall consensus is a value of $M_{500}$ roughly in the (1--2)$\E{14}\Msun$ range, consistent with our own estimates.

\begin{table}
  \begin{center}
    \caption{Mass Estimates}
    \label{tab:masses}
    \vspace{1ex}
    \begin{tabular}{clc}
      \hline
      Proxy & \multicolumn{1}{c}{Scaling relation} & $M_{500}$ $(10^{14}\Msun)$ \\
      \hline
      $Y$ & \secref~\ref{sec:a10mass} & $1.34\pm0.11$ \\
      \Lx & \secref~\ref{sec:xprior} & $1.0\pm0.2$\vspace{1ex}\\
      $Y$ & \citet{Andersson1006.3068} & 1.4 \\
      $Y$ & \citet{Planck1303.5080} & 1.7\vspace{1ex}\\
      \Lx & \citet{Pratt0809.3784} & 2.0 \\
      \Lx & \citet{Vikhlinin0805.2207} & 1.7 \\
      \Lx & \citet{Mantz0909.3099} & 1.1 \\
      \Lx & \citet{Andersson1006.3068} & 1.3 \\
      \hline
    \end{tabular}
  \end{center}
  \tablecomments{The evolution of cluster scaling relations out to $z=1.9$ has not been constrained by data, and so we have not attempted to quantify the systematic uncertainties on these estimates due to extrapolation in redshift. For the estimates obtained in \secref{}s~\ref{sec:a10mass} and \ref{sec:xprior}, the table reproduces the uncertainties quoted in each analysis; see those subsections for a discussion of exactly what is included in the error bars.}
\end{table}

Any cosmological interpretation of the detection of \cl{} should account for the survey selection function and take place in the context of appropriately calibrated scaling relations at $z>1$ \citep{Allen1103.4829}. However, we note that the mass estimates here, taken at face value, are entirely consistent with predictions of the concordance cosmological model \citep{Harrison1210.4369}.\footnote{Specifically, we ran the Matlab code provided by \citet{Harrison1210.4369} assuming a flat \LCDM{} model with $h=0.7$, $\Omegam=0.3$ and matter power spectrum amplitude $\sigma_8=0.8$. The code calculates three measures of rareness for a cluster with a given mass and redshift under a given cosmological model. Using a nominal mass and redshift of $M_{500}=1.34\E{14}\Msun$ and $z=1.91$, all three measures indicate that the existence of \cl{} within a 25 sq.\ deg.\ search area is consistent with the assumed cosmology. This conclusion does not change if the search area is 11 sq.\ deg.\ (corresponding to the precursor XMM-LSS survey).}

\subsection{SZ--X-ray Scaling} \label{sec:LYscaling}

\begin{figure}
  \centering
  \includegraphics{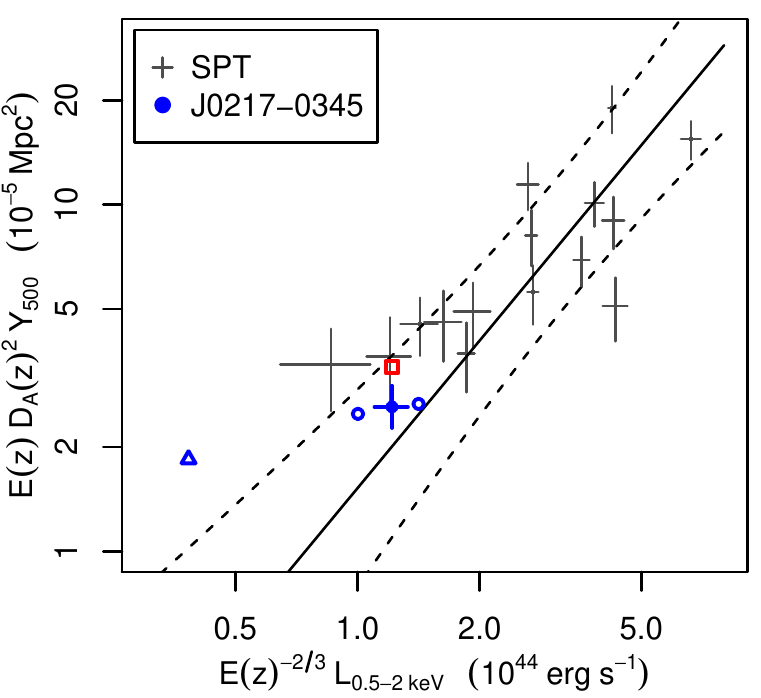}
  \caption{
    X-ray luminosity and SZ effect measurements for \cl{} are compared with measurements of SPT clusters from \citet{Andersson1006.3068}. Solid and dashed lines indicate the nominal $\Lx$--$Y_{500}$ relation fit to the SPT data and its 68.3\% confidence predictive range (\secref~\ref{sec:LYscaling}; the predictive range accounts for intrinsic scatter and uncertainty in the slope of the scaling relation, but not uncertainty in its evolution). The filled circle showing \cl{} is from our analysis in \secref~\ref{sec:xprior}, assuming a cluster redshift $z=1.91$. Open circles show the nominal results assuming $z=1.7$ and $2.1$, the limits of the likely redshift range identified from photometric data (\willis{}). The triangle shows our measurements if the cluster redshift is assumed to be 1.0, where the photometric data show a local maximum in the distribution of galaxy redshifts (although these galaxies are not clustered about the X-ray emission). The combination of X-ray flux and SZ data also disfavors a cluster redshift $z=1$ compared with $1.9$, assuming that the adopted evolution of the scaling relation holds out to $z\sim2$. The red square shows the best-fitting value, assuming $z=1.91$, when the pressure profile template of \citet{Planck-Collaboration1207.4061} is used rather than that of \sayers{}; the two agree at the $\sim1\sigma$ level (see \appref~\ref{sec:templates}).
  }
  \label{fig:LY_spt}
\end{figure}

\figref~\ref{fig:LY_spt} compares the \Lx{} and $Y_{500}$ values for \cl{} with measurements of SPT clusters (spanning $0.3<z<1.08$) from \citet[][hereafter \Aeleven{}]{Andersson1006.3068}. Those authors did not explicitly fit the \Lx{}--$Y_{500}$ scaling relation, but an approximate \Lx{}--$Y_{500}$ relation for the SPT clusters can be constructed by algebraically combining their results for the \Lx{}--$M_{500}$ and $Y_{500}$--$M_{500}$ relations. Including a unit conversion from keV$\Msun$ to Mpc$^2$, this is
\begin{equation} \label{eq:YLscaling}
  \frac{E(z) \dA(z)^2 Y_{500}}{10^{-5}\Mpc^2} = 1.48 \left( \frac{E(z)^{-2/3} \Lx}{10^{44}\erg\second^{-1}} \right)^{1.42}.
\end{equation}
Here, for consistency, we have followed \Aeleven{} in assuming a small departure from self-similar evolution in the \Lx--$M_{500}$ relation (ultimately based on the work of \citealt{Vikhlinin0805.2207}), leading to the combination $E^{-2/3}\Lx$ rather than $E^{-1}\Lx$. \eqnref~\ref{eq:YLscaling} appears as a solid line in \figref~\ref{fig:LY_spt}, and qualitatively fits the SPT clusters well.  At the smallest luminosities shown, bias from the SPT SZ selection will become increasingly important; the same bias prevents us from straightforwardly fitting the published \Aeleven{} measurements. Note that the \Aeleven{} scaling relation fits did approximately account for this selection bias.

Instead, we fit a power-law plus scatter model to the $\Lx/E^{2/3} > 2\E{44} \erg\second^{-1}$ SPT data, with the slope fixed to 1.42, finding an intrinsic scatter of $42^{+18}_{-12}$\%, marginally larger than the expected $\sim30$\% (based on a \Lx--$M$ scatter of 40--45\%; e.g., \citealt{Mantz0909.3099}). Incorporating this 42\% intrinsic scatter, and adopting an uncertainty of $\pm0.39$ on the slope, based on \Aeleven{}, yields the 68.3\% confidence  predictive range indicated by the dashed lines in \figref~\ref{fig:LY_spt}. With the assumed evolution, \cl{} is consistent with the SPT scaling relation. Adopting self-similar evolution would reduce this consistency, although not by a significant amount compared with the intrinsic scatter in the relation. Future work including a large sample of XXL clusters followed up with CARMA, covering a range of redshifts and luminosities, will provide better insight into the evolution of the scaling relation and related astrophysics.

The placement of \cl{} on this scaling diagram is dependent on the cluster redshift. While there is good evidence for a redshift $\sim1.9$ from photometric data (\willis{}), this has not yet been spectroscopically confirmed. \secref~\ref{sec:xprior} showed that our $Y_{500}$ constraint has a relatively mild dependence on the assumed redshift. In contrast, the rest-frame \Lx{} value inferred from the measured X-ray flux straightforwardly depends on redshift through factors of the luminosity distance and, to a lesser extent, the K-correction. Open circles in \figref~\ref{fig:LY_spt} show the nominal positions of \cl{} in the scaling-relation plot corresponding to cluster redshifts of $1.7$ and $2.1$, the limits of the likely redshift range identified by \willis{}; the differences from $z=1.91$ are small compared to the measurement errors and intrinsic scatter, and do not change our conclusions above.

Also shown in the figure (the triangular symbol) is the result assuming $z=1.0$, corresponding to a secondary peak in the distribution of galaxy redshifts from \willis{}. While the photometric data disfavor this redshift in any case (the $z\sim1$ galaxies are not clustered around the X-ray emission), \figref~\ref{fig:LY_spt} illustrates that requiring X-ray flux and SZ data to be consistent with a well calibrated scaling relation can in principle provide complementary information on galaxy cluster redshifts. If we assume that the adopted evolution of the scaling relation holds precisely, this consistency requirement favors a redshift $z=1.9$ over $z=1$ by a factor of $\sim5$, accounting for the measurement uncertainties, the intrinsic scatter, and the statistical uncertainty in the normalization and slope of the scaling relation. In practice, data such as those we present here (with the addition of spectroscopic redshift confirmation) should be used to better constrain the slope and evolution of the $\Lx$--$Y_{500}$ scaling relation, allowing this technique to be applied to new clusters discovered in the future.

Within the measurement uncertainties and intrinsic scatter, our conclusions in this section are robust against the particular pressure profile template (\sayers) adopted in this work. The red square in \figref~\ref{fig:LY_spt} shows the results that would be obtained (assuming $z=1.91)$ when adopting instead the template profile of the \citet{Planck-Collaboration1207.4061}. See further discussion in \appref~\ref{sec:templates}.

\section{Summary} \label{sec:conclusion}

We have detected the Sunyaev-Zel'dovich effect of XXL galaxy cluster \cl{}, confirming the presence of a hot ICM. With a photometrically determined redshift of $z=1.91^{+0.19}_{-0.21}$ (\willis{}), \cl{} is the most distant cluster for which the SZ effect has been detected. Extrapolating a variety of locally calibrated scaling relations, we estimate a mass in the range $M_{500}\sim (1$--$2)\E{14}\Msun$ from the X-ray and SZ data, with the caveat that such estimates are heavily dependent on the assumed evolution of the scaling relations. In contrast, our measurement of the  spherically integrated Comptonization, $Y_{500}=(3.0\pm0.4)\E{-12}$, is relatively insensitive to the cluster's redshift over a broad range.

For redshifts consistent with the photometric data, the measured $Y_{500}$ and X-ray luminosity are in good agreement with the extrapolation of a $Y_{500}$--$\Lx$ scaling relation calibrated from higher-mass and lower-redshift SPT clusters (\Aeleven{}). In principle, requiring consistency with such a scaling relation provides a way to constrain cluster redshifts based only on X-ray flux and SZ measurements. At the moment, uncertainties in the slope and especially in the evolution of the $Y_{500}$--\Lx{} relation are large, limiting the utility of this approach for relatively low-mass, high-redshift clusters such as \cl{}.  However, given a scaling relation that has been calibrated at similar masses and at redshifts $z>1$, SZ follow-up could provide confirmation and redshift and mass information for the population of low signal-to-noise, extended X-ray detections expected in new surveys such as eROSITA. 

We note that the SZ measurements presented here were made using only the CARMA sub-array of eight 3.5\,m antennas operating at a wavelength of 1\,cm. Using the new, low-noise cm-wave receivers recently installed on all 23 antennas, similar detection signal-to-noise could be achieved in roughly one tenth of the time, while also providing sensitivity over a larger range of angular scales.

\section*{Acknowledgments}

XXL is an international project based around an XMM Very Large Programme surveying two 25 sq.\ deg.\ extragalactic fields at a depth of $\sim5\E{-15}\erg\cm^{-2}\second^{-1}$ in the 0.5--2.0\keV{} band. The XXL website is \url{http://irfu.cea.fr/xxl}. Multi-band information and spectroscopic follow-up of the X-ray sources are obtained through a number of survey programs, summarized at\\ \url{http://xxlmultiwave.pbworks.com/}.

Support for CARMA construction was derived from the states of California, Illinois, and Maryland, the James S. McDonnell Foundation, the Gordon and Betty Moore Foundation, the Kenneth T. and Eileen L. Norris Foundation, the University of Chicago, the Associates of the California Institute of Technology, and the National Science Foundation (NSF). Ongoing CARMA development and operations are supported by the National Science Foundation under a cooperative agreement, and by the CARMA partner universities; the work at Chicago was supported by NSF grant AST-1140019. Additional support was provided by PHY-0114422. FP acknowledges support from BMBF/DLR grant 50 OR 1117 and the DfG Transregio Programme TR33.

{\it Facilities:} CARMA, XMM (EPIC)

\def \aap {A\&A} % alternative A&A code
\def \aapr {A\&AR} % alternative A&AR
\def \aaps {A\&AS} % alternative A&AS
\def \statisci {Statis. Sci.} % statistical science
\def \physrep {Phys. Rep.} % physical review
\def \pre {Phys.\ Rev.\ E} % physical review E
\def \sjos {Scand. J. Statis.} % Scandinavian Journal of Statistics
\def \jrssb {J. Roy. Statist. Soc. B} % Journal of the Royal Statistical Society. Series B (Statistical Methodology)
\def \pan {Phys. Atom. Nucl.} % Physics of Atomic Nuclei
\def \epja {Eur. Phys. J. A} % European Physical Journal A
\def \epjc {Eur. Phys. J. C} % European Physical Journal C
\def \jcap {J. Cosmology Astropart. Phys.} % Journal of Cosmology and Astro-Particle Physics
\def \ijmpd {Int.\ J.\ Mod.\ Phys.\ D} % international journal of modern physics D
\def \nar {New Astron. Rev.} % new astronomy review
\def \araa {ARA\&A}
\def \aj {AJ}
\def \aar {A\&AR}
\def \apj {ApJ}
\def \apjl {ApJL}
\def \apjs {ApJS}
\def \asl {Adv. Sci. Lett.} % Advanced Science Letters
\def \mnras {MNRAS}
\def \nat {Nat}
\def \pasj {PASJ}
\def \pasp {PASP}
\def \science {Sci}
\def \gca {Geochim.\ Cosmochim.\ Acta}
\def \npa {Nucl.\ Phys.\ A}
\def \plb {Phys.\ Lett.\ B}
\def \prc {Phys.\ Rev.\ C}
\def \prd {Phys.\ Rev.\ D}
\def \prl {Phys.\ Rev.\ Lett.}

\appendix

\section{Influence of the Pressure Profile Template} \label{sec:templates}

The CARMA sub-array of eight 3.5\,m telescopes sample a limited range of angular scales around $\sim1$ arcmin that are well matched to detecting the bulk SZ effect of distant clusters, but not to measuring the detailed shape of cluster pressure profiles. Consequently, some assumption about the shape of the profile must be made in order to reconstruct the three-dimensional Comptonization of a galaxy cluster from these data.

The parametrized form of the scaled pressure profile in \eqnref~\ref{eq:gnfw} was proposed by \citet{Nagai0703661}, and its parameters have been constrained from the combination of X-ray data and hydrodynamical simulations \citep{Arnaud0910.1234}, X-ray and SZ data \citep{Planck-Collaboration1207.4061}, and SZ data alone (\sayers{}). In this work, we adopt the \sayers{} template because it has the advantage of being fit to SZ data at all radii (rather than X-ray data at small radii and/or simulations at large radii). In this appendix, we compare results obtained employing the \sayers{} template to those which assume templates from \citet{Arnaud0910.1234} and the \citet{Planck-Collaboration1207.4061}, in order to estimate the impact of this choice.

\begin{figure}
  \centering
  \includegraphics{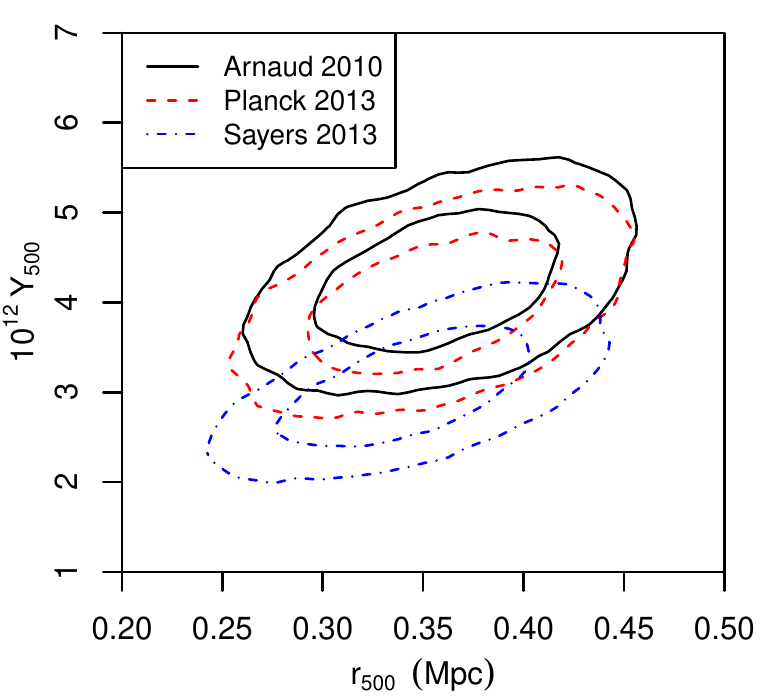}
  \caption{
    Two-dimensional 68.3 and 95.4\% confidence regions from an analysis as in \secref~\ref{sec:xprior}, comparing results employing the pressure profile templates given by \citet{Arnaud0910.1234}, the \citet{Planck-Collaboration1207.4061} and \sayers{}. Although the constraints are mutually consistent, the difference between the template profiles (primarily in the slope at large radii) does have an effect on our results.
  }
  \label{fig:templates}
\end{figure}

\figref~\ref{fig:templates} shows constraints from the analysis described in \secref~\ref{sec:xprior}, employing each of the pressure templates. (These are displayed in terms of $r_{500}$ rather than \rs{} because the templates specify different values of $c_{500}$; hence, our X-ray prior on $r_{500}$ translates to different values of the scale radius in each case.) The primary difference between the templates is the slope of the pressure profile at large radii, which is steepest for the \citet{Arnaud0910.1234} template and shallowest for the \sayers{} template (see \sayers{}). This is reflected straightforwardly in a relatively smaller value of $Y_{500}$ obtained from the \sayers{} template, since a larger fraction of the projected signal is ascribed to pressure at radii $>r_{500}$. We note, however, that the two-dimensional constraints shown remain consistent at the $1\sigma$ level. The marginalized constraints on $10^{12}Y_{500}$ are respectively $4.2\pm0.5$ (for the \citealt{Arnaud0910.1234} template), $4.0\pm0.5$ (for \citealt{Planck-Collaboration1207.4061}) and $3.0\pm0.4$ (for \sayers{}). In terms of \figref~\ref{fig:LY_spt}, the values derived using all three templates lie within the $1\sigma$ predictive interval of the SPT scaling relation, i.e., \cl{} is within the intrinsic scatter of the extrapolated relation regardless of which template is used. Note that the significance of the cluster detection, in terms of the improvement in $\chi^2$ when a model for the cluster is included in the fit, is similar across all three templates.

We next consider whether the $\ltsim1\sigma$ discrepancy between the constraints in \figref~\ref{fig:templates} might be reduced if uncertainty in each of the template shapes is taken into account. For the \citet{Arnaud0910.1234} template there is effectively no measurement error, since the profile at radii $r>r_{500}$ is assumed to match simulations.\footnote{\citet{Marrone1107.5115} showed that varying the shape of the \citet{Arnaud0910.1234} template at small radii had only a percent effect on $Y_{500}$ values derived from CARMA data.\\~\\~\\} In contrast, both the \citet{Planck-Collaboration1207.4061} and \sayers{} templates are based on SZ measurements of the pressure at large radii. We were able to obtain the full posterior distribution of the parameters determining the model in \eqnref~\ref{eq:gnfw} from the \sayers{} analysis (J.~Sayers, private communication). Marginalizing over this multi-dimensional prior expands the \sayers-template constraints to be consistent with the other templates at the $1\sigma$ level, as would be expected from the $\sim1\sigma$ consistency of the templates themselves within these uncertainties (\sayers). We cannot repeat this exercise for the \citet{Planck-Collaboration1207.4061} template, since the measurement uncertainties in that case are not available to us. However, it seems reasonable to conclude that the small discrepancy between these two templates can be explained by measurement uncertainties affecting the templates themselves.

Further study of the pressure in cluster outskirts, at and beyond the virial radius, would help to settle this question. We note that all of the works considered above include relatively strong priors by virtue of adopting a parametrized model (\eqnref~\ref{eq:gnfw}). A non-parametric approach such as the Gaussian process model investigated by \sayers{} may provide a better alternative, given sufficiently precise data; that particular option has the benefit of naturally including both measurement uncertainties and intrinsic scatter among clusters as a function of radius.

\end{document}